\documentstyle[12pt,aaspp4] {article}

\begin{document}

\title {A double peaked pulse profile observed \\in GX 1+4}
\author{ B. Paul, P. C. Agrawal, A. R. Rao and R. K. Manchanda}
\affil{Tata Institute of Fundamental Research\\Mumbai(Bombay), India, 400005}
\centerline {September 1996}
\centerline{ To appear in }
\centerline{ Astronomy and Astrophysics}

\begin{abstract}
The hard X-ray pulsar GX 1+4 was observed several times in the last few years
with a pair of balloon-borne Xenon filled  Multi-cell Proportional Counters
(XMPC). In a balloon flight made on 22 March 1995, the source was detected 
in a  bright state, the average observed source count  rate being  
$8.0\pm0.2$ s$^{-1}$ per detector. X-ray pulsations with a period of 
$121.9\pm0.1$ s were detected in the source with a broad double peak pulse 
feature. When observed in December 1993 with the same instrument, the 
pulse profile of GX 1+4 showed a single peak. This change in the pulse 
profile to a double pulse structure in about 15 months indicates 
either activation of the opposite pole of the neutron 
star if the magnetic field is asymmetric or possibly a change in the beam 
pattern, from a pencil beam to a fan beam. 
Assuming a fan beam configuration, the pulse profile is used to find the 
inclinations of the magnetic axis and the viewing axis with the spin axis. The 
derived angles 
support the {\it GINGA} observations of a dip in the pulse profile which was 
resolved to have a local maximum in one of the observations and was explained 
with resonance scattering of cyclotron line energy photons by the accretion column 
(Makishima et al., \markcite{maki1988}, Dotani et al., \markcite{dotani1989}.). Compared to our previous 
observation of the same source with the same telescope (Rao et al., \markcite{rao1994}) a 
period change rate of $0.72 \pm 0.40$ s yr$^{-1}$ is obtained which is the lowest rate of
change of period for this source since its discovery.

Average pulse fraction in the hard X-ray range is low $(30\%)$, 
consistent with its anti correlation with luminosity as reported by 
us earlier (Rao et al., \markcite{rao1994}) 
and the observed spectrum is very hard (power law photon index  $1.67\pm0.12$).

\end{abstract}

\section{Introduction}
The galactic center low-mass hard X-ray pulsar GX 1+4 was discovered
by Lewin et al. \markcite{lewin1971} in a balloon observation with a pulse period of 135 
seconds. The source X-ray luminosity with an assumed distance of 10 kpc, was 
estimated to be 10$^{37}$ $-$ 10$^{38}$ erg s$^{-1}$, very close to 
$L_{edd}$ for a small polar cap region.
Follow up observations confirmed the pulsations and showed a very 
fast spin-up rate with a time scale of 40 years. In the conventional accretion 
disc theory such a fast spin up was difficult to explain. In the early 80's 
the source made a transition to a low 
intensity state and attempts to detect the source with EXOSAT failed. The source 
was observable again in 1987 with the {\it GINGA} satellite and by then the 
spin change had reversed from spin-up to spin-down. Since then, until a very 
recent increase in luminosity, all the observations gave the same spin down trend 
with $\dot{P} = 1.5$ s yr$^{-1}$. The spin-up and the spin-down episodes of GX 1+4 
are explained by the disc accretion model developed by Ghosh and Lamb \markcite{ghosh1979a};\markcite{ghosh1979b}). 
The total torque acting on the central object is resultant of three torques, 
1) the torque carried by the in-falling matter, 
2) positive torque acting through the magnetic lines by the disc inside the 
radius where the Keplerian frequency is the same as the rotation frequency of 
the core object and 
3) negative torque acted through the magnetic lines by the disc outside the 
corotation radius. In the low luminosity state the third component can be 
dominant over the first two and a resultant negative torque will cause 
spin-down of the pulsar.

To improve the understanding of the neutron star magnetic 
field with an accretion disc and to verify the models of accretion torque 
on the neutron stars, periodic observations of GX 1+4 with a balloon-borne large 
area hard X-ray telescope were started in 1991. In four observations made so far 
the source was found in a high state during the last two times and the pulse 
period was 
determined accurately. Additionally, study of
the X-ray spectrum and pulse phased spectroscopy was also 
carried out. These results will be reported in detail in a later publication. Any observation of change 
in the energy spectrum with the pulse phase will give interesting information about 
either the emission regions or the environment through which photons come 
out in different phases of its 122 s spin period. In this paper we 
report our results on a change in the hard X-ray pulse profile and its 
implications in terms of the current accretion models.

\section{Observations}

The X-ray pulsar GX 1+4 was observed with a large area 
xenon-filled multi-anode proportional counter (XMPC) telescope in the 20--100 keV 
band in a balloon flight experiment on 22 March 1995. The balloon reached
a ceiling altitude of 41.5 km corresponding to a residual atmosphere 
of 2.5 gm cm$^{-2}$.

The XMPC telescope consists of two identical xenon-filled proportional 
counters with a total effective area of 2400 cm$^2$ and has a 5$^{\circ}$ $\times$ 5$^{\circ}$
field of view defined by a passive tin-copper graded
collimator. The telescope, mounted on an orientation platform, can be
pre-programmed to track a given source by an onboard automated tracking 
system. For details of the X-ray telescope refer to Rao et al. \markcite{rao1991}. 
Observations of the source were carried out by alternately looking at the 
source and a nearby source-free background region in the tracking mode. 
Background observations were also made in a tracking mode in a region 
about 8$^{\circ}$ away from 
the source and free of any other known bright X-ray sources.

During the 
first and the last 10 minutes of the 240 minutes observations of GX 1+4,
some noise was present in the lower energy channels in one of the proportional 
counters. All data from that detector during that time were discarded. Data 
in those lower channels for the entire duration of the observation were also 
discarded to remove any ambiguity. The background count rate was found to be 
constant during the entire balloon flight; the source count 
rate varied with the zenith distance because of absorption along the varying
residual atmosphere along the line of sight. The constancy of the 
background count rate was checked by fitting a straight line and a reduced 
$\chi^{2}$ of  1.1 was obtained. 

\section{Analysis and results }

\subsection { Timing analysis }

In each of the two detectors the 20-100 keV photons were detected with
128 channel energy information 
with a time resolution of 1.28 ms. The count rate was binned with 5 sec and 10 
sec bin widths and a period search was done in the 50-200 sec range with
an FFT algorithm based on the Lomb-Scargle method. For both the detectors very 
clear periodograms 
with single sharp peaks around 121.9 seconds were obtained. The reduced data 
length and energy range in one of the detectors gave a periodogram peak of 
smaller height compared to that in the other detector. A period search in two different 
broad energy bands also gave clear periodograms with smaller peaks at the same 
value of the period. Finally to improve accuracy in determination of the period,
data from both 
the detectors were added and a periodogram was obtained. The pulse period of 
GX 1+4 as seen on 22nd March 1995 is determined to be $121.88 \pm 0.09$ s. The 
false alarm probability of the 121.88 s peak in the periodogram for an 
average background rate and the number of data points used, was 
calculated to be negligible$(\sim 10^{-9})$. Pulsations in the same 
source were also detected in a previous balloon observation with the same 
telescope. The pulse period as seen on 11 December 1993 was $121.0 \pm 0.4$ s 
(Rao et al., \markcite{rao1994}. Over this period of 15 months, the overall spin down rate 
of $0.72 \pm 0.40$ s yr$^{-1}$ is somewhat smaller than the average spin down rate of 
1.4 s yr$^{-1}$ since 1987. 
{\it BATSE} observations in the intervening period have reported a reversal 
of the spin change rate (Chakrabarty et al., \markcite{chak1994}), from spin-down to spin-up
thereby supporting the smaller rate of change derived from the present
observation. Pulse profiles in different energy bands were obtained by folding
the photon 
counting rates with the measured period of 121.88 s. A plot of the pulse profile in 
the 20--100 keV energy range is shown in fig 1. for two cycles.
The pulse profiles were obtained by adding data from the two detectors. The pulse fraction 
in the 20--50 keV range is estimated to be $25\%$ and that in the 50--100 keV range it is $34\%$.
The anti correlation between the pulse fraction and the luminosity found in the 1993
observation is still found to exist. The 20--50 keV pulse profile, which is 
the most clear one, shows a wide pulse with a valley at the center or 
two pulses with unequal separation. A double peaked pulse profile similar 
in structure but narrower in width was seen earlier by Makishima et al. \markcite{maki1988} 
in the 2--20 keV range. In our earlier observation in December 1993 there 
was no indication of a double pulse and the detected pulse was also narrower. 
It is possible that during the recent source brightening, there might have 
been a gradual change in the emission, from a pencil beam to a fan beam, 
which is more common to a pulsar in its bright state. A phase difference in 
the pulse profile with the energy was explained by a switch over in the beam pattern 
for a cylinder of emission at higher luminosity (White et al., \markcite{white1983}.  To 
investigate whether the difference in the pulse fraction in the two energy bands
is significant, we have obtained the hardness ratios (ratio of counts in 
the 20--35 keV range to that in the 20--100 keV range)  in the pulsed
(phase 0.45 to 1.05 in the pulse profile in the top panel of {\it fig}. 1.) and 
unpulsed (phase 0.05 to 0.45 in the pulse profile) parts of the profile. The derived values  are $0.69\pm0.02$ and
$0.71\pm0.03$, respectively for the pulsed and the unpulsed part of the
profile. Hence we conclude that there  is no clear indication of any 
change in the pulse fraction with the energy. A detailed analysis of the spectra 
with the pulse phase has also been done and will be reported separately.

\subsection { Spectral analysis }
The observed energy spectrum was fitted well with an incident power law spectrum 
with a photon index $1.67 \pm 0.11$ (reduced $\chi{^2}$ = 1.1). A thermal
Bremsstrahlung model gave a 
temperature of 99 keV. Spectral fits were also attempted with two Compton
scattered Bremsstrahlung models.
A temperature of 18.5 keV and an optical depth of 7.7 was obtained
for the first model while the second model gave a temperature of 
17.5 keV and an optical depth of 6.8.
Pulse phased spectra for the pulsed and the non-pulsed components of the 122 s
spin period were also obtained. These spectra were also fitted well with 
the power law and the thermal Bremsstrahlung models with similar values of the 
parameters but somewhat larger error bars.  The  X-ray luminosity in the 20--100 keV 
range is deduced to be 2.5$\pm$0.3 $\times$10$^{37}$ erg s$^{-1}$ for  a distance of 
10 kpc with a $20\%$ uncertainty on the higher side.

\subsection {Pulse profile modeling}

Very complex changes in the pulse 
profile with luminosity are seen in many X-ray pulsars. An
intensity-dependent widely varying pulse profile was observed in the 
transient pulsar EXO 2030+375 which was modeled with both the fan and the 
pencil beams of unequal intensity from the two offset magnetic poles, the most 
complex modeling of a X-ray pulsar profile done so far (Parmar et al., \markcite{1989}).
The pulse profile observed in the high luminosity state changed as the source
strength dropped by a factor of 100 and in a later bright state the initial
bright state pulse profile was again seen. In EXO 
2030+375 the relative luminosity of the two poles was found to change by a 
factor of 10. A change by a factor of $10^2$ in the overall luminosity
and dominance fan beam emission over pencil beam emission was found when 
luminosity was $> 10^{37}$ erg s$^{-1}$. At lower luminosity ($< 10^{37}$ erg s$^{-1}$)
the emitting material is in the form of a slab over the polar cap and since it emits
more along the local field lines, this results in a pencil beam pattern. At the higher
accretion rate, the material goes closer to the pole before it is halted and
it is held more like a cylinder. In this case the emission is more in the direction
of the magnetic equator, resulting in a fan beam pattern.  

The observed change in the GX 1+4 pulse profile from December 1993 to March 1995 
can be explained in two ways. One possibility is an activation of the second pole, which is possible if the
magnetic field is asymmetric in latitude (so that the distribution of mass
accretion onto the two poles depends on the Alfven radius $r_{A}$ or in turn
on the luminosity). The second plausible explanation is a gradual change in the 
beam pattern, from a pencil beam to a fan beam in spite of a decrease in luminosity
by a factor of 3 in 20--100 keV energy band.
In our modeling we have assumed a simple fan beam pattern of GX 1+4 with 
a symmetric magnetic dipole and equal intensity on both sides of the equator 
with a constant overall emission. The luminosity is maximum towards 
the magnetic equator from the neutron star center and decays exponentially 
towards the poles. The sum of the two angles, $\theta_m$ the angle between the 
magnetic axis and the 
spin axis and $\theta_r$ the angle between the observer line of sight and the 
spin axis needs to be more than $\pi \over 2$ so that the line of sight 
crosses the magnetic equator twice in one period and shows two peaks. 
Intensity has two minima, corresponding to the phases when the two poles 
are closest to the viewing axis. Such simple considerations were used 
successfully to reproduce roughly the pulse profiles of many pulsars by 
Leahy \markcite{leahy1991} \markcite{leahy1990}. To get the detailed features of pulse 
profiles, many other possibilities like offset in the two magnetic poles, 
unequal brightness of the two sides, gravitational bending near the neutron 
star surface for photons direction not normal to the surface, unequal size of 
the two emission regions etc. are to be considered. But for a pulse profile with 
few bins and relatively large errors on the data points, a simple geometry as 
described above gave reasonably good fit and we obtained the following values 
for the parameters
\newline
$ {\theta_m} = 56 \pm 8,~~~{\theta_r} = 56 \pm 8$ with ${\theta_m} + 
{\theta_r} = 112 \pm 2 $
\newline
and the exponential intensity decay towards the pole has an angular scale of
$ {\theta_d} = 32 \pm 4 $.  

The model considered here is actually unable to distinguish between 
${\theta_m}$ and ${\theta_r}$ because of their interchangeability. 
However the values we have obtained are the same for both the parameters. The 
constraint is more on the sum of the two angles which defines the closest 
position of the second pole with the viewing axis
$(180^{\circ} - {\theta_m} - {\theta_r})$ and produces the valley in between 
the two peaks. Similar value of the two angles ${\theta_m}$ and ${\theta_r}$ 
ensures that we see very close to the first pole at phase 0.25 and the 
intensity there is an overall background emission.
Two {\it GINGA} observations in 1987 and 1988 in 10-37 keV range discovered
two peculiar pulse profiles (Dotani et al., \markcite{dotani1989}). In the first 
observation at the peak of the profile, there was a dip with a local maximum 
in it and the intensity was $ 1.2 \times 10^{36}$ erg s$^{-1}$. In the second 
observation about $150^{\circ}$ away from the peak, again there was a dip but 
without any local maximum there unlike the previous observation and the 
intensity was $ 5.8 \times 10^{36}$ erg s$^{-1}$.
A hollow cylinder of accretion column causing resonance scattering 
at the energy of the cyclotron line explained the first observation. 
At the center of the column there was no scattering and that resulted in the 
local maximum. At a higher intensity level in the 
second observation, the accretion column was full and the local maximum in the 
dip was absent. For this to happen the observer has to see just through one 
of the poles and that is supported by nearly the same values of 
${\theta_m}$ and ${\theta_r}$ that
we have obtained. The offset of the dip with the peak in the pulse 
profile as observed
in the second {\it GINGA} observation is also explained with the present 
value of ${\theta_m}$ and ${\theta_r}$. In the second observation probably 
a gradual change from fan beam to pencil beam was taking place with an 
increase in luminosity, and the peak in the second observation is at the 
place of the two magnetic equator crossings and the dip is at the phase when 
one is seeing through the first pole. A larger value of ${\theta_d}$ can 
produce the wide peak in the second observation and the valley also may become 
less significant. GX 1+4 showed both single and double peaked pulse profiles 
on different occasions (Mony et al., \markcite{mony1991}). We have
observed both types of pulse profiles on two different occasions with the
same X-ray telescope.

The source geometry obtained here with the double peaked pulse profile can 
generate the single peaked profile observed in 1993 December if a pencil 
beam emission is considered. Very regular observations of GX 1+4 and accurate 
measurement of luminosity, pulsation period, period derivatives and
epochs may help in 
establishing this scenario of change in the beaming pattern.

\section{Discussion}

A continuous observation of this pulsar in its present bright state may 
improve our understanding of the disk-magnetosphere interactions in neutron 
stars and will also lead to verification of the same. As the luminosity in the
present state is comparable to its luminosity in the earlier spin-up phase, a 
reversal of the spin change will help to determine the critical value of fastness 
parameter at which the resultant torque on the neutron star changes its sign. 
GX 1+4 being a hard X-ray object, most of its luminosity lies in the 20--100 
keV region. To study the $L_{X}$ Vs $\dot{P}$ relation regular hard X-ray 
observations are needed. In our two observations, the period values indicate a very small 
spin-down rate. Compared to our previous observation of the same source 
with the same telescope, a period change rate of $0.72 \pm 0.40$ is obtained 
which is the lowest rate of change of period for this source since its 
discovery. A reversal in the spin change, from spin-down to spin-up in between 
our two observations is consistent with our period determinations. {\it BATSE} 
observations (Chakrabarty et al., \markcite{chak1994}) in this period has also indicated 
the same.

In the recent 1995 observation the pulse profile is double peaked, 
which may be an indication of increased mass accretion on the other magnetic 
pole. A gradual change in the beam pattern from a pencil beam to a fan beam,
also may explain this pulse pattern.
We note that the fan beam appeared in the relatively lower luminosity
state contrary to the present understanding. Fan beam is more likely to 
occur in a luminosity state of $ L > 10^{37}$ erg s$^{-1}$ but in the 1993
December observation it showed a simple pencil beam pattern in spite of a very
high luminosity of $7 \times 10^{37}$ erg s$^{-1}$. 
The pencil beam in the December 1993 observation can be explained 
if the time scale of change in the beam pattern is long (100 days).
The anti correlation between the luminosity and the pulse 
fraction, that was noticed in an earlier observation is still persistent and 
it will be interesting to observe whether this feature is present in other 
pulsars. 

Compared to the earlier observations ( Laurent et al. \markcite{laurent1993}) the spectrum 
is much harder which is evident from a power law photon index of 1.67. 
The thermal Bremsstrahlung model also fits well with a temperature of 99 keV
which is higher than that measured from the earlier high energy observations.

Significant difference between the pulsed and unpulsed spectra was not observed. 
To detect any small difference in the hardness or temperature, longer duration 
observations or much reduced background level and bigger effective area 
telescope will be required. 

A simple modeling of the double peaked pulse 
profile gave similar values for the inclination angles for the magnetic axis 
and the line of sight with the spin axis of the neutron star. The geometrical 
description that fits the present observation also supports the explanation 
given by Dotani et al. \markcite{dotani1989} for the dip structure seen in two {\it GINGA } 
observations. Our earlier observation of the source profile with single peak 
also can be explained if a pencil beam emission is considered.
One may also note that sometimes in fan beam 
configuration a larger value of the angular decay scale may hide the valley in 
between the two peaks corresponding to the two magnetic equator crossings. More 
frequent long duration observation of the source to determine accurate value 
of the luminosity, period, period derivative and epoch will help in establishing this 
possibility of change in the beam pattern.

\acknowledgements

It is a pleasure to acknowledge the contribution of Shri M.R. Shah, 
Electronics Engineer-in-charge of this experiment. We are also thankful 
to Shri D.K. Dedhia,  Shri K. Mukherjee,  Shri V.M. Gujar, Shri S.S.
Mohite, Shri P.B. Shah and Shri D. M. Pawar for their support in the fabrication of the
payload. We thank the Balloon Support Instrumentation
Group and the Balloon Flight Group led by Shri M.N. Joshi, 
under the overall supervision of Prof. S.V. Damle.

\begin{figure} 
\vspace*{200mm}
\includegraphics{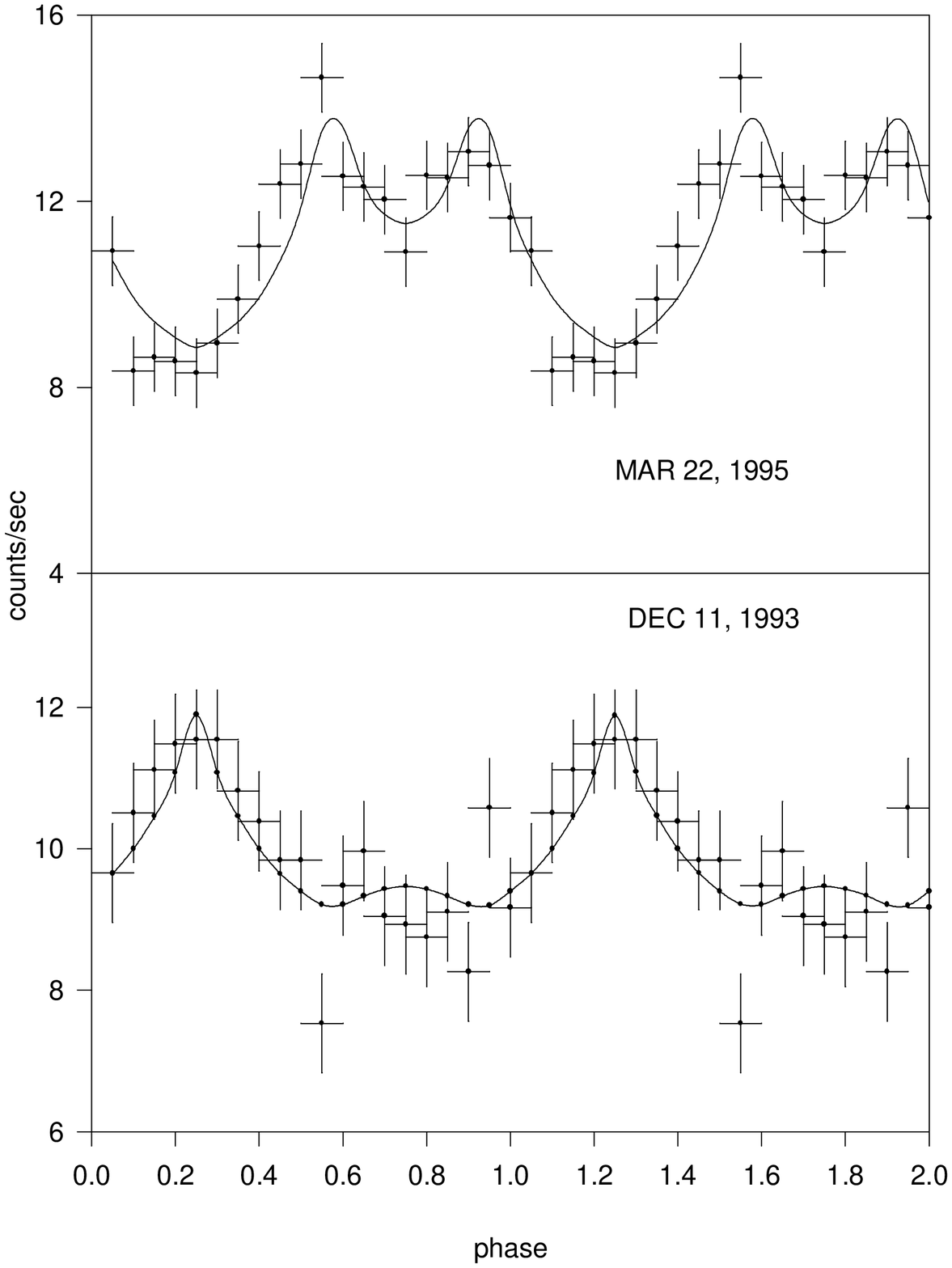} 
\caption
{Pulse profile of GX 1+4 obtained from the XMPC observations,
on two different occasions, plotted in two cycles for clarity.
Data from the two detectors have
been added to reduce the error in each bin. The lines represent the fan beam 
and the pencil beam emission patterns in the two cases as shown in the figure.
The phase alignment in the two observations is done arbitrarily with the
assumption that the pulse profile simulation discussed in the text is valid.}
\end{figure} 
\end{document}